\documentclass[11pt]{article}
\usepackage{textcomp}
\usepackage{pdfsync}
\usepackage{fancyhdr}
\usepackage{amssymb}
\usepackage{amsmath}

\usepackage{graphicx}
\usepackage{latexsym}
\usepackage{appendix}
\usepackage{srcltx}
\def\v{\nu}

\def\centeron#1#2{{\setbox0=\hbox{#1}\setbox1=\hbox{#2}\ifdim
   \wd1>\wd0\kern.48\wd1\kern-.48\wd0\fi
   \copy0\kern-.48\wd0\kern-.48\wd1\copy1\ifdim\wd0>\wd1
   \kern.48\wd0\kern-.48\wd1\fi}}

\def\PRL{Phys. Rev. Lett.~}
\def\PR {Phys. Rev.~}

\newcommand{\beq}{\begin{equation}}
\newcommand{\eeq}{\end{equation}}
\newcommand{\bea}{\begin{eqnarray}}
\newcommand{\eea}{\end{eqnarray}}
\newcommand{\ba}{\begin{array}}
\newcommand{\ea}{\end{array}}

\newcommand{\nn}{\nonumber}

\newcommand{\half}{\frac{1}{2}}

\begin{document}

\hskip3cm

 \hskip12cm{CQUeST-2013-0626}
\vskip3cm

\begin{center}
 \LARGE \bf    Thermodynamics from field equations for black holes with multiple horizons
\end{center}

\vskip2cm

\centerline{\Large Yongjoon Kwon\footnote{ykwon@sogang.ac.kr}~~{\rm and}~~Soonkeon Nam
\footnote{nam@khu.ac.kr}
}
\hskip2cm

\begin{quote}
Department of Physics and Research Institute of Basic Science, Kyung
Hee University, Seoul 130-701, Korea$^{1,2}$

Center for Quantum Spacetime, Sogang University, Seoul 121-741,
Korea$^1$
\end{quote}

\hskip2cm

\vskip2cm

\centerline{\bf Abstract}  The first law of black hole thermodynamics can be read off from the field equations at the horizon. Until now, for black holes with multiple horizons  the field equations only at the outer horizon were employed with a particular constraint. In this paper, however, we suggest that for a black hole with multiple horizons  the field equations at the inner horizon as well as the outer horizon should be needed in order to obtain  the first law of black hole thermodynamics in general.
  \thispagestyle{empty}
\renewcommand{\thefootnote}{\arabic{footnote}}
\setcounter{footnote}{0}
\newpage

\section{Introduction}

It has been well known that the black hole has thermodynamic properties such as the Hawking temperature and the entropy from the Bekenstein-Hawking area law, and these quantities satisfy the first law of black hole thermodynamics, $TdS=dE$.  About a decade ago, Padmanabhan showed that the Einstein field equations can be interpreted as a thermodynamic identity \cite{Padmanabhan:2002sha}. For this, the temperature at the horizon is  defined from the notion of the periodicity in Euclidean time, so that the temperature at the horizon($r=a$) is proportional to the differentiation of the metric component at the horizon, $f'(r=a)$, in the spherically symmetric metric, $ds^2=-f(r)dt^2+f^{-1}(r)dr^2+r^2d\Omega^2$ \cite{Padmanabhan:2002sha}. Then, the relevant tensor component of the field equations in the Einstein gravity, i.e., $G^{r}_{~r}= 8\pi G T^{r}_{~r}$, can be expressed as the first law of black hole thermodynamics.
After few years, it was found that the field equations in the Gauss-Bonnet gravity give the thermodynamic first law as well \cite{Paranjape:2006ca}. There have been also a lot of investigations on checking if the statement still holds for the black holes in any other gravity theories \cite{Kothawala:2007em,Akbar:2007qg,Akbar:2006mq,Cai:2009ph,Padmanabhan,Multicollect,Mirza:2011fp,Son:2013eea}. In particular, in the work of \cite{Akbar:2007qg} the rotating BTZ black hole in the Einstein gravity has been considered and obtained the thermodynamic first law from the relevant tensor of the field equations as follows:
\beq
-dE+TdS+\Omega dJ +P_r dA =0 \,,
\eeq
where $E$ is the mass of the black hole, $T$ is the Hawking temperature at the horizon, $S$ is the black hole entropy, $\Omega$ is the angular velocity at the horizon, $J$ is the angular momentum, $P_r= T^{r}_{~r}$ is the radial pressure from the source, and $dA$ is the change in horizon area. The term $P_r dA$ corresponds to work done against the pressure. Indeed, since the rotating BTZ black hole is the solution in Einstein field equations without no source term, $P_r$ should vanish in the above equation. 

For the rotating BTZ black hole, there are two horizons, of which one is the outer(event) horizon and the other is the inner horizon, because of the two black hole parameters, mass and angular momentum. It means that they can be written in terms of the both horizons. Therefore when we take the differentials of the black hole parameters, the variations of the inner horizon as well as the outer horizon should be considered. However, to obtain the above thermodynamic first law in \cite{Akbar:2007qg}, the only variation of the outer horizon was considered by assuming a particular variation of the two horizons.  
Moreover, in other papers about the thermodynamic interpretation of the field equations for black holes with multiple horizons only the variation of the outer horizon was considered under some similar particular variations \cite{Kothawala:2007em,Multicollect}.
In this paper, however, we will consider the general variations of both horizons without any particular constraint in order to find the first law of black hole thermodynamics.

When we consider a black hole, generically the  black hole can have three black hole parameters such as mass, angular momentum and electric(or magnetic) charge which are conserved quantities. In general, the thermodynamic first law of a black hole with three black hole parameters is given by
\beq
-dE+TdS+\Omega dJ+\Phi dQ =0 \,,
\eeq
where $\Phi$ is the potential from the source, and $Q$  is the conserved charge for the source.
According to Wald's formalism, the entropy can be also regarded as a conserved quantity \cite{wald,Kim:2013zha}. 
While the temperature($T$) and angular velocity($\Omega$) only depend on a given metric, the conserved quantities such as black hole mass($E$), entropy($S$), angular momentum($J$) and charge($Q$) given by a source  depend on gravity theory. 

In this paper, we will focus on the three dimensional gravity with negative cosmological constant. Particularly, the rotating BTZ black hole with two horizons will be considered as an example. As mentioned above, in the previous work \cite{Akbar:2007qg}, the rotating BTZ black hole in the Einstein gravity has been considered and  the thermodynamic first law was checked from the field equations at the outer horizon. However, in \cite{Akbar:2007qg} a particular variation of two horizons was implicitly assumed. By considering the angular velocity at the horizon as a constant for a variation, the variation of the inner horizon was related to the one  of the outer horizon as $dr_-={{r_-} \over {r_+}}dr_+$. Therefore, only field equations at the outer horizon could be used to obtain the thermodynamic first law.
However, we suggest that if a black hole has multiple horizons, the field equations corresponding to each horizon should be considered on the same footing in general. By considering the rotating BTZ black holes in the higher derivative gravity theories as well as the Einstein gravity, we will show that the relevant tensors of the field equations at both horizons give the first law of black hole thermodynamics under the general situation.

\section{Thermodynamics from the field equations at the horizon}
In this section, we will briefly review on the Padmanabhans' work \cite{Padmanabhan:2002sha,Paranjape:2006ca} that the relevant tensor of the field equations at the horizon can be written as the first law of  black hole thermodynamics. 
The action in the four dimensional Einstein gravity is given by
\beq
{\cal S}= {1 \over {16 \pi G}} \int d^{4} x \sqrt{-g}  R + {\cal S}_{{\rm matter}} \,.
\eeq
Via the variation of the action with respect to the metric $g_{\mu \nu}$, the Einstein field equations are obtained as
\beq
G_{\mu \nu}= 8 \pi G T_{\mu \nu} \,,
\eeq
where $G_{\mu \nu} \equiv R_{\mu \nu} -{1 \over 2} g_{\mu \nu} R$, and the energy momentum tensor from the matter part of the action is given by $T_{\mu \nu}= -2 (\sqrt{-g})^{-1}  { {\delta {\cal S}}_{{\rm matter}}} / {\delta g^{\mu \nu}} $.
When a black hole solution of the field equations has the single horizon at $r=a$ and the metric form is given by
\beq
ds^2=-f(r) dt^2+ f(r)^{-1} dr^2+r^2 d\Omega^2_{2} \,,
\eeq
the relevant tensor of the field equations is given as follows:
\beq
\label{reltensor}
G^{r}_{~r}={{ -1+f(r)+r {f'(r)}} \over  r^2} =   8 \pi G P \,,
\eeq
where ${P =  T^{r}_{~r}}$ means the radial pressure of the source.
Note that  the Hawking temperature at the horizon $(r=a)$ is related to $ T= f'(a) / 4 \pi $.
Multiplying by the differential volume $dV= 4 \pi a^2 da$ on both sides of the equation (\ref{reltensor}) at the horizon, the field equation becomes
\beq
4\pi ({{4\pi a T -1}  }) da = 8 \pi G P dV \,,
\eeq
where $f(a)=0$ is used.
Then, it is found that this implies the  first law of the thermodynamics as follows:
\beq
{{T dS - dE }  }  = P dV \,,
\eeq
identifying with the entropy, $S={ {\pi a^2} \over {G}}={{A_H} \over {4G}}$, and the energy, $E= {a \over {2 G}}= \big( {{A_H} \over {16\pi G^2}} \big)^{1/2}$. 
The $PdV$  term represents the work done by the radial pressure.
If there is no matter(source) field, the radial pressure provided by the source is absent, i.e., $P=0$.

In the next section, we will consider the rotating BTZ black hole with multiple horizons as a three dimensional  black hole solution in  the Einstein gravity and the higher derivative gravity theories such as topologically massive gravity and new massive gravity. For three dimensional case, the unit $c=8G=\hbar=1$ is used through this paper.

\section{BTZ black holes in three dimensional gravity theories}

\subsection{Einstein gravity}

First, let us consider the Einstein gravity with negative cosmological constant whose action is given by 
\beq
{{\cal S}_{EH}}= {1 \over {16 \pi G}} \int d^{3} x \sqrt{-g} \Big( R + {2 \over \ell^2} \Big)  + {\cal S}_{{\rm matter}} \,,
\eeq
and, through the variation of the action, the field equations are obtained as 
\beq
\label{BTZEOM}
G_{\mu \nu}-{1 \over \ell^2}g_{\mu\nu}=\pi  T_{\mu \nu}  \,,
\eeq
where $T_{\mu \nu}$ is the energy momentum tensor for the matter(source) field.
When there is negative cosmological constant, it contributes to the energy of the spacetime as background vacuum energy. 
To find the first law of black hole thermodynamics written in terms of the differentials of the black hole's conserved quantities, we would rather not include the cosmological term into a matter field action in the following, due to the result of the work \cite{Son:2013eea} about the thermodynamics from the field equations for the AdS black hole in the Gauss-bonnet gravity. 
The rotating BTZ black hole has the metric form of 
\beq
\label{BTZmetric}
ds^2=-f(r) dt^2+ f(r)^{-1} dr^2+r^2 \left(d\phi- {J \over {2 r^2}} dt \right)^2 \,, 
\eeq
where $f(r) =-M+{r^2 \over \ell^2}+{J^2 \over {4 r^2}}= { 1 \over {r^2 \ell^2}} {(r^2-{r^2_+})(r^2-{r^2_-})}$  \cite{Banados}.
As seen, there are two black hole parameters which generate two horizons. Since for the rotating case the differential $dJ$ as well as the differential $dM$ should be taken into account, we should consider the variations of both horizons, $dr_-$ and $dr_+$, respectively.
 
From the relevant tensors of the field equations at the outer horizon $(r=r_+)$ and the inner horizon ($r=r_-$), we obtain the following two equations by multiplying with the differential volume $dV=2 \pi r dr$ at each horizon on the both sides:
\begin{eqnarray}
\label{EinBTZA}
\bigg(-{{2 {r_+}} \over \ell^2}+ { {2 {r^2_-}} \over {r_+  \ell^2}}+{ {f'({r_+})} } \bigg) d{r_+} =  P_+ dV_+  =0 \,, \\  
\label{EinBTZB}
 \bigg( -{{2 {r_-}} \over \ell^2}+{ {2 {r^2_+}} \over {{r_-}  \ell^2}} +{ {f'({r_-})} } \bigg) d{r_-} =  P_- dV_- =0 \,, 
\end{eqnarray}
where $P_{\pm}=T^{r}_{~r}(r={r_{\pm}})$ are the radial pressures at the outer and inner horizons.
Note that since the rotating BTZ black hole is a black hole solution in the Einstein gravity with no matter(source), the radial pressures vanishes, i.e., $P_{\pm}=0$.

By summing the two equations, (\ref{EinBTZA}) and (\ref{EinBTZB}), we obtain the following equation: 
\beq
\label{sumEinBTZ}
-{ {2 } \over \ell^2} \big( {r_+} d{r_+} +{r_-} d{r_-} \big) + { {2 } \over \ell^2} \bigg({ {{r^2_-}} \over {{r_+} }}d{r_+} +  { { {r^2_+}} \over {{r_-} }}d{r_-} \bigg) +4\pi \big(T_+ d{r_+} - T_- d{r_-} \big) = \big(P_+ dV_+ + P_- dV_- \big) =0 \,,
\eeq
where $T_+=f'({r_+})/4 \pi$ and $T_-=-f'({r_-})/4 \pi$. Here, since $f'({r_-})$ is negative we take the temperature at the inner horizon   as  positive one.
Using the eq.(\ref{EinBTZB}), it is easily found that the equation (\ref{sumEinBTZ}) implies the thermodynamic first law of the BTZ black hole as 
\beq
-dM+\Omega_+ dJ + T_+ dS =0 \,,
\eeq
with the  conserved quantities,
\beq
 M = {{r^2_+ +{r^2_-}} \over \ell^2}  \,,~~~ J= {{2 {r_+} {r_-}} \over \ell }  \,,~~~ S= 4 \pi {r_+} \,,
\eeq
and the angular velocity at the outer horizon, $\Omega_+= {{r_-} / {({r_+} \ell)}}$.
%

\subsection{Topologically massive gravity}

In this section, we will consider the topologically massive gravity(TMG) as one of higher derivative gravity theories, which also has the BTZ black hole solution.  
The action of TMG consists of the Einstein action with negative cosmological constant and gravitational Chern-Simons term as follows \cite{Deser}:
\begin{eqnarray} \label{action}
{ {\cal S}_{TMG}} &=& {{\cal S}_{EH} } + {{\cal S}_{CS} } \\
  &=& {1 \over {16 \pi G}} \int  d^3 x { \sqrt{-g}  \left( R+{ 2 \over {l^2} } \right)} 
- {l \over {96 \pi G \v}} \int  d^3 x  {\sqrt{-g}   \varepsilon ^{\lambda \mu \nu} \Gamma ^{r} _ {\lambda \sigma} \left( \partial_\mu \Gamma ^{\sigma} _{r \nu} +{2 \over 3}  \Gamma ^{\sigma} _{\mu \tau}  \Gamma ^{\tau} _{\nu r} \right)} ,  \nonumber 
\end{eqnarray}
where $\v$ is the coupling constant and  $\varepsilon ^{012}=+1 / \sqrt{-g} $ is the Levi-Civita tensor. 
Varying this action with respect to the metric, the field equations are obtained as
\begin{equation} \label{revieq}
G_{\mu \nu} -{1 \over l^2}   g_{\mu \nu} + {l \over  {3 \v} }  C_{\mu \nu} =0 \,,
\end{equation}
where  $C_{\mu \nu} $ is the Cotton tensor given by
\begin{eqnarray}
 C_{\mu \nu} =  \varepsilon  _{\mu} ^{~\alpha \beta}    \nabla _\alpha \left( R_{\beta \nu} -{1 \over 4}  g_{\beta \nu}  R \right) \,.
\end{eqnarray}

With the rotating BTZ metric (\ref{BTZmetric}), it is found that the relevant tensors of the field equations at both horizons give the following equations:
\begin{eqnarray}
\label{TMGBTZA}
\bigg(1+ {{r_-} \over  {3 {r_+} \v}} \bigg) \bigg (-{2 {r_+} \over \ell^2}+ { { 2 {r^2_-}} \over {{r_+} \ell^2}}  +{f'(r_+)}  \bigg ) dr_+ =  P_+ dV_+ =0 \,,\\
\label{TMGBTZB}
\bigg(1+ {{r_+ } \over  {3 {r_-} \v}} \bigg) \bigg (-{2 {r_-} \over \ell^2}+ { {2 {r^2_+}} \over {{r_-} \ell^2}}  + {f'(r_-)}  \bigg )dr_- =   P_- dV_- =0 \,,
\end{eqnarray}
where the both sides are multiplied by the differential volume $dV=2 \pi r dr$ at each horizon.
When $\v \rightarrow \infty$, it can be easily seen that these two equations reduce to the eqs.(\ref{EinBTZA}) and (\ref{EinBTZB}) in the Einstein gravity case.
Likewise, using the eq.(\ref{TMGBTZB}), the sum of the above two equations can be written as follows:
\beq
\label{TMGtheq}
\big(-dM+ \Omega_+ dJ+ T_+ dS \big)+{1 \over {3 \nu}} \Bigg[-{1 \over \ell}dJ +\Omega_+ \ell dM+ T_+ \left({{S} \over {r_+}}d{r_-}-{{S {r_-}} \over {r^2_+}}dr_+ +{{r_-} \over {r_+}}  dS \right)  \Bigg] =0 \,.
\eeq
Note that $M$, $J$, and $S$ correspond  to the conserved charges in the Einstein gravity case. While the angular velocity  $\Omega_+$ and the temperature $T_+$ only depend on the given metric regardless of gravity theories,  the conserved charges in TMG have some changes because of the higher derivative terms. Therefore the mass, angular momentum and entropy in TMG are different from the Einstein gravity case as follows \cite{Bouchareb:2007yx,Solodukhin:2005ah,Deser:2005jf,Kim:2013cor}: 
\beq
{\cal M}= M+{J \over {3 \v  \ell}} \,,~~~ {\cal J}= J+{{M \ell} \over {3 \v}} \,,~~~ {\cal S} =\Big(1+{{r_-} \over {3 \v {r_+}}} \Big) S \,.
\eeq
In terms of these quantities, it is easily seen that the eq.(\ref{TMGtheq}) can be written as 
\beq
-d{\cal M}+\Omega_+ d{\cal J}+T_+ d{\cal S} =0 \,.
\eeq
This is just the same as  the thermodynamic first law for the rotating BTZ black hole in TMG.
%


\subsection{New massive gravity}

It is also known that the rotating BTZ black hole is a solution in another higher derivative gravity theory,  the so-called new massive gravity(NMG) whose action is given by \cite{Bergshoeff}
\begin{eqnarray} 
\label{NMGaction}
{\cal S}_{NMG}={1 \over {16 \pi G }}\int d^3x\sqrt{-g}\bigg[ R +{2 \over \ell^2} + \frac{1}{m^2} \Big(R_{\mu\nu}R^{\mu\nu} -\frac{3}{8}R^2 \Big)  \bigg]  \,.
\end{eqnarray}
Through the metric variation, the field equations are obtained as
\begin{equation}
\label{NMGEOM}
   G_{\mu\nu} - {1 \over \ell^2} g_{\mu\nu} + \frac{1}{2m^2}K_{\mu\nu} 
        =0\,,
\end{equation}
 where
\begin{equation}
 K_{\mu\nu} = g_{\mu\nu}\Big(3R_{\alpha\beta}R^{\alpha\beta}-\frac{13}{8}R^2\Big)
                + \frac{9}{2}RR_{\mu\nu} -8R_{\mu\alpha}R^{\alpha}_{\nu}
                + \half\Big(4\nabla^2R_{\mu\nu}-\nabla_{\mu}\nabla_{\nu}R
                -g_{\mu\nu}\nabla^2R\Big) \nn\,.
\end{equation}

For the rotating BTZ black hole, we will consider the same metric form as eq.(\ref{BTZmetric}) with different AdS radius $L$, so that $f(r) =-M+{r^2 \over  L^2}+{J^2 \over {4 r^2}}= { 1 \over {r^2 L^2}} {(r^2-{r^2_+})(r^2-{r^2_-})}$ .
Note that in order to be a solution in NMG the following relation between $L$ and $\ell$ should be satisfied:
\bea \label{schrel}
{1 \over \ell^2}= {1 \over L^2} \bigg(1- {1 \over {4 m^2 L^2}} \bigg) \,.
\eea
With this relation, it is found that the relevant tensors of the field equations at both horizons are given by
\begin{eqnarray}
\label{NMGBTZA}
\bigg[ 1-{1 \over {2 m^2 L^2}} \left( 1-{{6 {r^2_-}} \over {r^2_+}} \right) \bigg] \bigg( -{{2 {r_+}} \over L^2}+{ {2 {r^2_-}} \over {{r_+} L^2}} +f'(r_+) \bigg) dr_+ = P_+ dV_+ =0  \,, \\
\label{NMGBTZB}
\bigg[ 1-{1 \over {2 m^2 L^2}} \left( 1-{{6 {r^2_+}} \over {r^2_-}} \right) \bigg] \bigg( -{{2 {r_-}} \over L^2}+{ {2 {r^2_+}} \over {{r_-} L^2} } +f'(r_-) \bigg) dr_- = P_- dV_-=0 \,,
\end{eqnarray}
where the differential volume $dV=2 \pi r dr$ at each horizon is multiplied on the both sides.
 As the limit of $m \rightarrow \infty$ is taken, these two equations reduce to the eqs.(\ref{EinBTZA}) and (\ref{EinBTZB}) in the Einstein gravity.

Using the eq.(\ref{NMGBTZB}), the sum of the two equations can be written as follows:
\beq
\label{TMGeq}
\beta \big(-dM+\Omega_+ dJ+T_+ dS \big)= \big(P_+ dr_+ +P_- dr_- \big)=0 \,,
\eeq
where $\beta \equiv  \big(1- {1 \over {2 m^2 L^2}} + {{3 \Omega^2_+} \over m^2} \big) $. 
Note that $M$, $J$, and $S$ correspond to the conserved charges of BTZ black hole with AdS radius $L$ in the Einstein gravity.
While the angular velocity  $\Omega_+$ and the temperature $T_+$ are same with the Einstein gravity case, the conserved charges in NMG are slightly different on account of higher derivative terms as follows \cite{Clement:2009gq,Nam:2010ub,Kwon:2011jz,Kim:2013qra}:
\beq
{\tilde M}= \alpha M   \,,~~~ {\tilde J}= \alpha J \,,~~~ {\tilde S}= \alpha S \,,
\eeq
where $\alpha \equiv  \big(1+{1 \over {2 m^2 L^2 }} \big)$.
Therefore, in terms of these quantities, the eq.(\ref{TMGeq}) can be written as 
\beq
 -d{\tilde M}+\Omega_+ d{\tilde J}+T_+ d{\tilde S}  = {\alpha \over \beta}  \big(P_+ d{r_+}+P_- d{r_-} \big)=0 \,.
\eeq
This is the same as the thermodynamic first law for the rotating BTZ black hole in NMG.
Notice that in this case the proportionality constant is needed to read off the thermodynamic first law from the field equations.

\section{Conclusion}

Since the Padmanabhan's proposal that the first law of black hole thermodynamics can be directly read off from the relevant tensor of the field equations, there have been many investigations about the consistency of the statement by considering specific black holes in diverse gravity theories. Until now, for black holes with multiple horizons the thermodynamic first law has been induced from the field equations only at the outer(event) horizon with a constraint for variations. In the previous work for the rotating BTZ black hole in the Einstein gravity \cite{Akbar:2007qg}, we found that it has been done under the assumption of $d\Omega_+=0$. Even in other works, by considering a particular variation of the two horizons only the field equations at the outer horizon were employed \cite{Multicollect}.
In this paper, however, we have considered the general variations of the two horizons without any constraint and showed that when a black hole has multiple horizons, the field equations at the inner horizon as well as the outer horizon should be considered on the same footing. As the examples, the rotating BTZ black holes with two horizons were considered in the Einstein gravity and higher derivative gravity theories. 

As other examples of black holes with multiple horizons, the Reissner-Nordstr{$ {\rm \ddot o}$}m black hole in the Einstein gravity and new type black hole in new massive gravity were considered in the appendix and the thermodynamic first laws of those black holes were obtained by using the field equations at both horizons. 
Therefore it is now clearly understood why the field equations at both horizons should be used in the previous work \cite{Mirza:2011fp} for the new type black hole in NMG. Not only for the new type black hole in NMG but also for any black holes with multiple horizons the field equations at both horizons are needed. 

There was also the previous work on the thermodynamic first law of the Kerr-Newmann black hole from the field equations \cite{Kothawala:2007em}, in which only field equations at the outer horizon were used, even though the black hole has two horizons, by considering a particular variation, $dJ= a dM$. 
However, if we consider general differentials of the black hole parameters, the field equations at the inner horizon should be considered together with the ones at the outer horizon.

Contrary to black holes with the outer and inner horizons, the Schwarzschild de-Sitter black hole, which has the cosmological and black hole horizons, have little different aspects.  
The two horizons of the de-Sitter black hole occur because of the black hole mass and cosmological constant. However, when we consider the differentials of the black hole parameters in the first law of black hole thermodynamics, there is no differential of the cosmological constant since it is usually regarded as just a constant, not a dynamical variable. In other words, it is not considered as a black hole parameter. 
This means that the two horizons are not independent practically, so that only the field equations at only one horizon -it could be the black hole horizon or the cosmological horizon- are enough to get the thermodynamic first law at the one horizon.

\section*{Acknowledgements}


This work was supported by the National Research Foundation of
Korea(NRF) grant funded by the Korea government(MSIP) through the
Center for Quantum Spacetime (CQUeST) of Sogang University with grant
number 2005-0049409 and also supported by Basic Science
Research Program through the National Research Foundation of
Korea(NRF) funded by the Ministry of Education (No.2013R1A1A2004538).

\newpage
\appendix
\section{Reissner-Nordstr{$ { \rm \ddot o}$}m black hole in the Einstein gravity}
The field equations in the four dimensional spacetimes without cosmological constant are given as $G_{\mu \nu}=8 \pi G T_{\mu \nu}$. For  the Reissner-Nordstr{$ {\rm \ddot o}$}m black hole, the energy momentum tensor, $T_{\mu \nu}$, is not zero since there is a source.
The metric of the Reissner-Nordstr{$ {\rm \ddot o}$}m black hole has the form of
\beq
 \label{RNmetric}
ds^2=-f(r) dt^2+ f(r)^{-1} dr^2+r^2 (d\theta^2+ \sin^2{\theta} d\phi^2) \,,
\eeq
where $ f(r) = 1-{ {2 G M } \over r} +{{G Q^2} \over r^2} = {{(r-{r_+})(r-{r_-}) } \over r^2}$.

Now, let us consider the relevant tensors of the field equations at the inner horizon as well as the outer horizon.  By multiplying with the differential volume $dV=4 \pi r^2 dr$ at each horizon, it is obtained as 
\begin{eqnarray}
  \label{RNA}
 {1 \over {2 G}}\big( {4 \pi {r_+}}  T_+ - 1 \big) dr_+ = P_+ dV_+ \,, \\ 
  \label{RNB}
 -{1 \over {2 G}} \big( {{4 \pi {r_-}} } T_- +1 \big) dr_- = P_- dV_- \,, 
\end{eqnarray}
where $P_{ \pm}=T^{r}_{~r}(r_{\pm})=-{{Q^2} / {8 \pi r^4_{\pm}}} $, and $T_{\pm}= \pm f'(r_{\pm}) / 4 \pi >0$.

Using the eq.(\ref{RNB}), the sum of the two equations can be written as 
\beq
- {1 \over {2 G}} \big( dr_+ +dr_- \big)  + T_+ {{4 \pi r_+} \over {2 G}} dr_+   +{1 \over {2 G}} dr_- -{Q^2 \over {2 {r^2_-}}} dr_- =  - {Q^2 \over {2 {r^2_+}}}dr_+  - {Q^2 \over {2 {r^2_-}}}dr_-   \,.
\eeq 
Therefore it is found that this implies the first law of black hole thermodynamics,
\beq
 -dM+ T_+ dS + \Phi_+ dQ =0 \,,
\eeq
with the conserved quantities, 
\beq
M= {{r_+ +r_-} \over {2 G}}  \,, ~~~ S={{ \pi r^2_+} \over G} \,,~{\rm and}~~  Q^2 = {{r_+ r_-} \over  G}  \,,
\eeq
where $\Phi_+= {Q / r_+}$ is the electric potential at the outer horizon.

\section{New type black hole in the new massive gravity}

At the special point, $2 m^2 L^2=1$, the field equations (\ref{NMGEOM}) have the new type black hole as well as BTZ black hole as solutions \cite{Bergshoeff}. For simplicity, we will consider the (static) new type black hole with two horizons from the two parameters in the metric. 
Since the metric is given in the form of 
\beq
ds^2= -f(r)dt^2+f(r)^{-1}dr^2+r^2 d\phi^2 \,,
\eeq
where $f(r)= {r^2 \over L^2} +{{ b r} \over L} +c ={{(r-r_+)(r-r_-)} \over L^2}$ \,,
the relevant tensor of the field equations is obtained as
\beq
\label{NMGNEOM}
\left(1-{1 \over {2 m^2 L^2}} \right)\Bigg[ {-{1 \over L^2}+{{f'(r)} \over {2 r}}}\Bigg] = \pi P =0 \,.
\eeq
Note that each parameter in the metric is not a black hole parameter as the conserved quantity.
When $2 m^2 L^2 \neq 1$, the parameter $b$ should be zero to satisfy this equation, which corresponds to the non-rotating BTZ black holes when $c<0$. However, at the special point the second parenthesis can be nonzero. Indeed, this happens for the new type black hole. In this case the second parenthesis in the eq.(\ref{NMGNEOM}) at each horizon is given by
\begin{eqnarray}
\label{NMGNTA}
\Big( -{2 {r_+} \over L^2}  + {4\pi T_+} \Big)  dr_+= -  \Big({{r_+ +r_-} \over L^2} \Big) dr_+ \,, \\
\label{NMGNTB}
\Big( -{2 {r_-} \over L^2}  -{4\pi T_-}  \Big) dr_- = - \Big({{r_+ +r_-} \over L^2} \Big) dr_-  \,,
\end{eqnarray}
where $dV=2 \pi r dr$ at each horizon is multiplied on the both sides.
The right-hand side is from the existence of the parameter $b$ by the higer derivative terms which gives the weaker boundary conditions than the Brown-Henneaux boundary conditions \cite{Brown:1986nw}. Therefore the new type black hole has some different physical phenomena from black holes in the Einstein gravity.
Note that for the non-rotating BTZ black hole case at the special point, the right-hand side is zero since $r_+=-r_-$.

The sum of the two equations is written as
\beq
-{{r_+ -r_-} \over L^2} \big(dr_+ -dr_- \big) + 4 \pi \big(T_+ dr_+ - T_- dr_- \big)  =0 \,.
\eeq
Using the eq.(\ref{NMGNTB}), with the following conserved quantities for the new type black hole \cite{Nam:2010ub,Kwon:2011jz,Kim:2013qra,Kwon:2011ey},
\beq
M= {\big({r_+ - r_- }\big)^2 \over {2 L^2}} \,,  ~~~~{\rm and}~~~~    S = 4\pi \big(r_+ -r_- \big)\,,
\eeq 
we find the first law of thermodynamics for the new type black hole as follows:
\beq
-dM + T_+ dS  =0 \,.
\eeq

\newpage

\end{document}